\documentclass[aps,prl,twocolumn,groupedaddress,showpacs]{revtex4}
\usepackage{graphicx}

\begin{document}

\preprint{Submitted to Phys. Rev. Lett.}

\title{
Quantum Non-Locality in Systems with Open Boundaries: \\
Failure of the Wigner-Function Formalism
}


\author{Luigi Genovese$^{1,2}$}
%\email[]{}
%\thanks{}
\author{David Taj$^{1,3}$}
%\email[]{}
%\thanks{}
\author{Fausto Rossi$^{1}$}
\email[]{Fausto.Rossi@PoliTo.It}
%\thanks{}
\affiliation{
$^1$
Dipartimento di Fisica, Politecnico di Torino, 10129 Torino, Italy \\
$^2$
Dipartimento di Fisica e Sez. INFN, Universit\`a di Roma ``Tor Vergata'', 00133 Roma, Italy \\
$^3$
Dipartimento di Matematica, Universit\`a di Torino, 10123 Torino, Italy
}

\date{\today}

\begin{abstract}

We shall revisit the conventional treatment of open quantum devices based on the Wigner-Function formalism. Our analysis will show that the artificial spatial separation between device active region and external reservoirs ---properly defined within a semiclassical simulation scheme--- is intrinsically incompatible with the non-local character of quantum mechanics. More specifically, by means of an exactly-solvable semiconductor model, we shall show that the application of the conventional boundary-condition scheme to the Wigner transport equation may produce highly non-physical results, like thermal injection of coherent state superpositions and boundary-driven negative probability distributions.

\end{abstract}

%\keywords{}

\maketitle

%\section{}
%\subsection{}
%\subsubsection{}

% If in two-column mode, this environment will change to single-column
% format so that long equations can be displayed. Use
%\begin{widetext}
%\end{widetext}

Current micro/nanoelectronics technology pushes device dimensions
toward limits where the traditional semiclassical Boltzmann
theory~\cite{SBT} can no longer be applied, and more rigorous
quantum-transport approaches are required~\cite{QT}. However, in
spite of the genuine quantum-mechanical nature of carrier dynamics
in the active region of typical nanostructured devices ---like
semiconductor superlattices and double-barrier structures--- the
overall behavior of such quantum systems is often the result of a
non-trivial interplay between phase coherence and
dissipation/dephasing~\cite{Interplay}, the latter being also due
to the presence of spatial boundaries~\cite{WF}. It follows that a
proper treatment of such novel nanoscale devices requires a
theoretical modelling able to properly account for both coherent
and incoherent ---i.e., phase-breaking--- processes on the same
footing. To this aim, a commonly used theoretical instrument is
the so-called single-particle density matrix~\cite{DM,GF-note};
however, while the latter is the ideal tool for the description of
ultrafast phenomena in infinitely-extended/periodically-repeated
nanostructures~\cite{RMP}, it cannot be directly applied to
quantum systems with open boundaries, for which a real-space
treatment is imperative.

Such a real-space description is naturally provided by the
Wigner-function formalism~\cite{WF}; within this approach the
statistical quantum state of the electronic subsystem is fully
described in terms of the so-called Wigner function, a function
defined over the conventional phase-space as the Weyl-Wigner
transform of the single-particle density matrix \cite{RMP,PRL}.
For the case of a one-dimensional problem, the equation describing
the time evolution of the Wigner function ---often referred to as
the Wigner transport equation--- is of the form:
\begin{equation}\label{e:WE1}
\frac{d}{dt} f(z,k) = \int dz' dk'\,
{\cal L}(z,k; z',k') f(z',k') \ ,
\end{equation}
where ${\cal L}(z,k;z',k')$ denotes the corresponding effective
Liouville superoperator~\cite{PRL} written in the Weyl-Wigner
phase-space representation $z,k$.

Different approaches for the study of quan\-tum-tran\-sport
phe\-no\-me\-na in semi\-con\-duc\-tor nano\-struc\-tu\-res based
on the Wigner-function formalism have been proposed. On the one
hand, starting from the pioneering work by Frensley
\cite{Frensley}, a few groups~\cite{WF-appl} have performed
quantum-transport simulations based on a direct numerical solution
of the Wigner equation in (\ref{e:WE1}) via finite-difference
approaches by imposing to $f(z,k)$ the standard boundary-condition
scheme of the semiclassical device modelling, also called ${\bf
U}$-scheme~\cite{BR}. On the other hand, a generalization to
systems with open boundaries of the well-known Semiconductor Bloch
Equations (SBE)~\cite{DM,RMP} ---the set of equations governing
the time evolution of the single-particle density matrix--- has
been recently proposed~\cite{PRL,BR}; such generalized SBE
---obtained again via the ${\bf U}$-scheme previously mentioned---
describe the open nature of the problem via a boundary source term
and a corresponding renormalization of the Liouville
superoperator. In addition to the two alternative simulation
strategies previously recalled ---both based on effective
treatments of relevant interaction mechanisms---, Jacoboni and
co-workers have proposed a fully quantum-mechanical simulation
scheme for the study of electron-phonon interaction based on the
so-called ``Wigner paths''~\cite{Modena,Modena-note}.

The generalized SBE approach previously recalled has been recently
applied to prototypical semiconductor-based open systems.
Preliminary results presented in \cite{BR} seem to suggest an
intrinsic limitation of the conventional Wigner function formalism
in describing quantum-transport phenomena through systems with
open boundaries. On the other hand, no clear evidence of such
limitations has been reported so far via Wigner-function
simulations based on finite-difference
treatments~\cite{WF,WF-appl}. Aim of this Letter is to solve this
apparent contradiction, thus shedding light on the real
limitations of the conventional Wigner-function picture applied to
open-device modelling. Our analysis will show that {\em the
artificial spatial separation between device active region and
external reservoirs is intrinsically incompatible with the
non-local character of quantum mechanics}.

In order to gain more insight into this highly non-trivial
problem, let us start considering the explicit form of the Wigner
equation (\ref{e:WE1}) in stationary conditions and in the absence
of any scattering mechanism:
\begin{equation}\label{e:WE2}
v(k)\,\frac{\partial}{\partial z} f(z,k) + \int dk' {\cal
V}(z,k-k') f(z,k') = 0\ ,
\end{equation}
where
\begin{equation}\label{e:calV}
{\cal V}(z,k) = \int_{-\infty}^{+\infty}\hspace{-.2cm} dz'\; \frac{e^{-2 i k z'}}{i \pi \hbar}  \left[V\left(z+z'\right) - V\left(z-z'\right)\right]\,
%{e^{-i k z'} \over i 2 \pi \hbar}
\end{equation}
is the Weyl-Wigner superoperator corresponding to the device
potential profile $V(z)$, while $v(k)$ denotes the electron group
velocity. Following the standard ${\bf U}$-scheme, we shall now
impose the desired spatial boundary conditions for $f$ at the left
($z = -{l \over 2}$) and right ($z = +{l \over 2}$) contacts,
specifying the ``incoming'' electron distribution $f_b(k) =
f\left(z_b(k),k\right)$, where $z_b(k) = {l \over
2}\,\left(\theta(-k)-\theta(+k)\right)$ denote the left and right
boundary coordinate corresponding, respectively, to positive and
negative carrier wavevectors $k$ ($\theta$ being the usual
Heaviside step function). By integrating Eq.~(\ref{e:WE2}) from
the spatial boundary $z_b(k)$ to the current point $z$ we
get~\cite{Dyson-note}:
\begin{eqnarray}\label{e:fs1}
f(z,k) &=& f_b(k) - \int_{z_b(k)}^z \!\!\!dz' \int dk'\, {{\cal V}(z',k-k') \over v(k)} f(z',k')\nonumber\\
&=& f_b(k) + \int \!dz' dk'\;{\cal W}(z,k;z',k') f(z',k')
\end{eqnarray}
with
$
{\cal W}(z,k;z',k') = -\theta(z'-z_b(k))\theta(z-z'){{\cal V}(z',k-k') \over v(k)}
$.
In a compact notation we have:
\begin{equation}\label{e:fs2}
f = f_b + {\cal W} f \ ,\quad
{\rm or}\quad
f = {1 \over 1-{\cal W}}\, f_b \ .
\end{equation}
By expanding the above formal solution in powers of the
interaction superoperator/propagator ${\cal W}$ ---and therefore
of the potential ${\cal V}$--- we get the well-known Neumann
series:
\begin{equation}\label{e:NE}
f = \sum_{n=0}^\infty {\cal W}^n f_b \ .
\end{equation}

Let us now focus on the case of a symmetric potential profile
($V(z) = V(-z)$), which in turn corresponds to an antisymmetric
potential superoperator, i.e., ${\cal V}(z,k) = -{\cal V}(-z,k)$.
Using this property together with the symmetric nature of our
spatial boundaries, i.e., $z_b(k) = -z_b(-k)$, it is possible to
show that the interaction propagator ${\cal W}$ is also preserving
the potential symmetry~\cite{ab-note}. This result is extremely
important: it implies that for any symmetric potential profile and
spatial boundaries the Neumann expansion in (\ref{e:NE}) gives
always a Wigner function symmetric in the spatial coordinate:
$f(z,k) = f(-z,k)$. Therefore, in total agreement with the
numerical results of the generalized SBE presented in \cite{BR},
the spatial charge density $n(z) = \int f(z,k)\, dk$ is always
symmetric, no matter which is the shape of the injected carrier
distribution $f_b(k)$. As anticipated, such symmetric behavior
---which is an exact result of the treatment presented so far---
has never been observed via finite-difference calculations.

In order to solve this apparent contradiction, let us now focus on
a particular choice of the device potential profile: $ V(z) =
{V_{_0} \over 2} \left[1+\cos(2 \kappa z)\right] $. The
corresponding superoperator in (\ref{e:calV}) is simply given by:
\begin{equation}\label{e:calV-cos}
{\cal V}(z,k) =
\pi  V_{_0} \sin(2 \kappa z) [\delta(k-\kappa)-\delta(k+\kappa)]\ .
\end{equation}
For this particular potential super\-ope\-ra\-tor
---cha\-rac\-te\-ri\-zed by a factorization/decoupling of position ($z$) and momentum ($k$) coordinates---
it is possible to obtain the
spatial charge distribution analytically:
\begin{equation}\label{e:AR}
n(z) = \int \hspace{-.05cm} dk f_b(k)\, {_1F_2}\left[\frac{1}{2}; \, 1+\alpha(k),\, 1-\alpha(k);\, q(z)\right]
\end{equation}
with
\begin{equation}\label{e:q}
q(z) = \frac{2 m^* V_{_0}}{\kappa^2}\,
\sin[\kappa(z-\frac{l}{2})]\, \sin[\kappa(z+\frac{l}{2})]\ .
\end{equation}
Here, $m^*$ is the electron effective mass, $\alpha(k)=k/\, \kappa$, and ${_1F_2}$ denotes the generalized
hypergeometric function of type $(1,2)$.

\begin{figure}[h]
\includegraphics[width=6cm,angle=270]{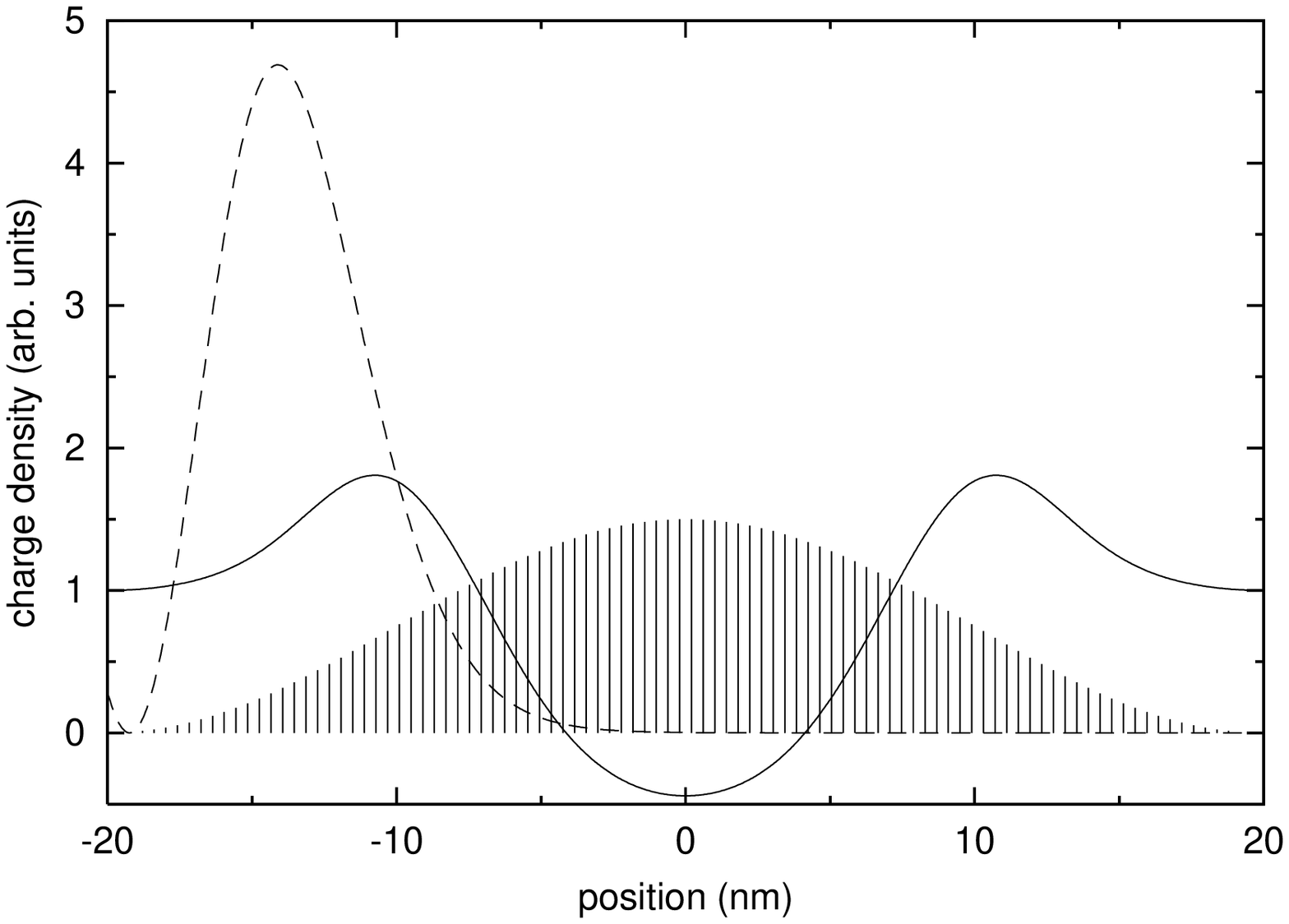}
\caption{Quantum-transport through a GaAs-based device active region ($l = 40$\,nm) with a cosine-like potential
profile ($V_{_0} = 150$\,meV) sandwiched between its electrical contacts: comparison between the analytical charge
distribution (solid curve) and the corresponding phenomenological result (dashed curve) for the case of a monoenergetic
carrier injection from left ($E_\circ = 70$\,meV). \label{fig1}}
%\end{figure}
%\begin{figure}
\includegraphics[width=6cm,angle=270]{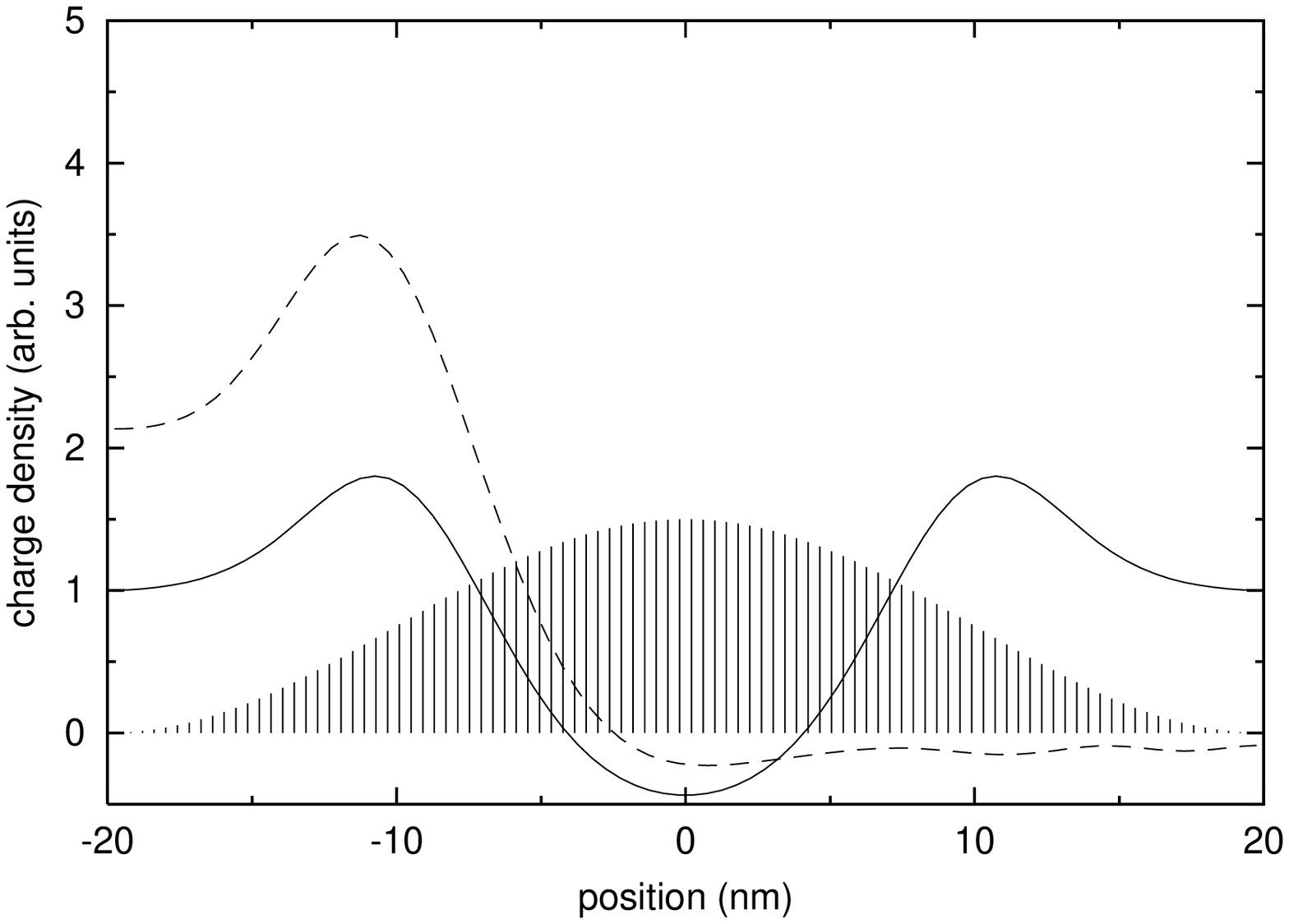}
\caption{Finite-difference result (dashed curve) and
a corresponding iterative numerical solution of Eq.~\ref{e:fs1} (see text). \label{fig2}}
\end{figure}

In order to investigate the main features of the analytical
results obtained so far, let us start by considering extremely
simple spatial boundary conditions: a monoenergetic carrier
injection from left only, i.e., $f_b(k) \propto
\delta(k-k_\circ)$. Moreover, we choose a potential periodicity
$\kappa = {\pi \over l}$ corresponding to just one maximum within
the device active region, so as to mimic the single-barrier device
considered in \cite{BR}. Figure \ref{fig1} shows a comparison
between the analytical spatial charge distribution in (\ref{e:AR})
and the phenomenological result obtained via conventional
scattering-state calculations. As we can see, the two curves
differ significantly: while the phenomenological charge
distribution (dashed curve)
---describing an extremely small tunnelling dynamics through the
potential maximum--- exhibits a clear and non-ambiguous asymmetric
behavior, the analytical result of the Wigner-function approach
(solid curve) is always symmetric, in total agreement with the
analysis proposed in~\cite{BR}. This implies that, also for an
infinitely high potential barrier and for a monoenergetic
injection from left only, carriers are ``instantaneously'' present
also on the right part of the device, a typical non-local feature
of our quantum-mechanical calculation. More important, within the
Wigner-function treatment (solid curve) we deal with a spatial
charge/probability distribution with negative values, which tells
us that {\it the analytical solution $f(z,k)$ of the differential
equation in (\ref{e:WE1}) is not necessarily a Wigner
function}~\cite{NV-note}.

Let us finally come to potential discrepancies between exact and
finite-difference results. Figure \ref{fig2} shows a comparison
between the conventional finite-difference  solution of
Eq.~(\ref{e:WE2}) proposed in~\cite{WF} (dashed curve) and a
numerical iterative solution of Eq.~(\ref{e:fs1}) (solid curve);
for both cases the same $80 \times 80$ phase-space discretization
has been employed. As we can see, while the iterative solution
coincides with the analytical result (see solid curve in
Fig.~\ref{fig1}), the finite-difference result comes out to be
strongly asymmetric. A closer inspection (not reported here)
reveals  that such anomalous behavior is mainly ascribed to the
usual non-symmetric discretization (left or right derivative) of
the kinetic/diffusion term in Eq.~(\ref{e:WE2}), which may result
in a fictitious decoherence/damping dynamics~\cite{IS-note}. This
seems to indicate that the more regular ---and physically-sound---
results obtained so far via finite-difference calculations may be
ascribed to such non-symmetric discretization procedure, which in
turn tends to limit the highly non-physical features of the
Wigner-function formalism applied to systems with open boundaries.

The analysis presented so far allows us to draw a few important
conclusions. First of all, the results of the analytically
solvable device model previously considered clearly show that the
usual boundary-condition scheme ---successfully applied to
semiclassical device modelling--- is intrinsically incompatible
with the non-local nature of quantum mechanics. More specifically,
from a strictly mathematical point of view it is true that the
Wigner equation (\ref{e:WE2}) is a first-order equation in $z$
parameterized by the wavevector $k$ and, as such, it is sufficient
to specify the value of the Wigner function at the device
boundaries; on the other hand, from a physical point of view the
separation between active region and external reservoirs/contacts
is only apparent, since the action of the potential superoperator
${\cal V}$ [see Eq.~(\ref{e:WE2})] is local in space, but its
value inside the device depends on the properties of the potential
$v(z)$ both inside and outside the device active region [see
Eq.~(\ref{e:calV})]. We are forced to conclude that the
application of the standard boundary-condition scheme to the
Wigner equation in (\ref{e:WE2}) is not physically justified,
since it may provide solutions which are not Wigner functions,
i.e., which do not correspond to the state of a quantum system. A
clear and unambiguous proof of such non-physical outcomes are the
negative values of the electron probability distribution reported
in Figs.~\ref{fig1} and \ref{fig2}. Generally speaking, what is
intrinsically wrong in the usual Wigner-function treatment of open
devices is the spatial separation between active region and
reservoirs. This is similar to isolate a portion of a given energy
spectrum, and try to treat such subset of energy levels as an
independent subsystem. In contrast, in order to provide a rigorous
description of open quantum devices it is imperative to introduce
a clear separation between the degrees of freedom of the subsystem
of interest (i.e., electrons primarily localized within the device
active region) and those of the external reservoirs (i.e.,
carriers primarily localized within the electrical contacts).
Given such separation, it is possible to obtain an effective
transport equation for the subsystem of interest via a suitable
statistical average over the coordinates of the ``environment''.
Following such reduction procedure, one gets again the Wigner
equation (\ref{e:WE1}), where now the Liouville superoperator is
the sum of a contribution describing the subsystem/device of
interest plus an effective superoperator describing the
interaction of the device with external reservoirs: ${\cal L} =
{\cal L}^{dev} + {\cal L}^{dev-res}$. We stress that within the
proposed alternative approach there is no need to impose any
spatial boundary condition: as for the case of a closed system the
spatial domain is again $-\infty < z < +\infty$. The only boundary
condition is the Wigner function $f(z,k)$ at the initial time
$t_\circ$; it follows that ---contrary to the case of spatial
boundary conditions--- the homogeneous Liouville dynamics in
(\ref{e:WE1}) provides/maintains a ``good'' Wigner function at any
later time $t$, thus preventing from non-physical behaviors, like
those reported in Fig.~\ref{fig1}.

We finally stress that the alternative simulation scheme described
so far may be concretely realized following the prescription
recently proposed in \cite{CSP}: the basic idea is to replace the
usual open description of quantum devices based on spatial
boundaries (i.e., source/loss terms) with a closed-system
treatment, where the interaction of the device electrons with the
external reservoirs is simply described in terms of effective
scattering rates acting on the device electrons only.


\begin{thebibliography}{xx}
\bibitem{GF-note}
It is worth mentioning that an equivalent and alternative approach
to the density-matrix formalism is provided by the nonequilibrium
Green's function technique~\cite{HJ}. Following this approach,
Lake, Datta, and co-workers have proposed an efficient
quantum-transport simulation scheme for the study of open
semiconductor devices~\cite{GF-appl}.
\bibitem{Modena-note}
This approach is intrinsically able to overcome the standard
approximations of the semiclassical scattering theory; however,
due to the huge amount of computation required, its applicability
is often limited to short time-scales and extremely simplified
situations.
\bibitem{Dyson-note}
The above formal solution can be also regarded as a sort of
Schwinger-Dyson equation linking the desired Wigner function $f$
to the free-particle one ($f_\circ(z,k) = f_b(k)$) through the
potential superoperator ${\cal V}$.
\bibitem{ab-note}
More specifically, by applying the interaction superoperator ${\cal W}$ to any symmetric function ($a(z,k) = a(-z,k)$),
we get as a result again a symmetric function of $z$, i.e.,
\begin{eqnarray}
b(z,k) &=& -\int_{z_b(k)}^z dz' \int dk' {{\cal V}(z',k-k')\over v(k)} a(z',k')\nonumber
\\ &=& b(-z,k).
\label{e:ab}
\end{eqnarray}
\bibitem{NV-note}
Indeed, it is well known that a ``good'' Wigner function, i.e., obtained as the Weyl-Wigner transform of a density
matrix, will never give rise to negative charge distributions.
\bibitem{IS-note}
We stress that this problem does not affect the iterative solution of Eq.~(\ref{e:fs1}), since in this case the kinetic
part of the dynamics is treated exactly.
\end{thebibliography}
\end{document}